# Viscosity-model-independent generalized Reynolds number for laminar flow of shear-thinning and viscoplastic fluids

Coskun Bilgi, and Niema M Pahlevan

[1] Department of Aerospace and Mechanical Engineering, University of Southern California, Los Angeles, CA, United States

[2] Division of Cardiovascular Medicine, Keck School of Medicine, University of Southern California, Los Angeles, CA, United States

Understanding the flow dynamics of non-Newtonian fluids is crucial in various engineering, industrial, and biomedical applications. However, the existing generalized Reynolds number formulations for non-Newtonian fluids have limited applicability due to their dependencies on their specific viscosity models. In this study, we propose a new viscosity-model-independent formulation for a generalized Reynolds number. The proposed method is based on the direct adaptation of the measurement principles of rotational viscometers for wall shear rate estimation. We assess the accuracy of this proposed formulation for power-law and Carreau-Yasuda viscosity models through robust friction factor experiments. The experimental results demonstrate the applicability and effectiveness of the proposed viscosity-model-independent Reynolds number, as the measured friction factor data align closely with our Reynolds number predictions. Furthermore, we compare the accuracy of our Reynolds number formulation against established generalized Reynolds formulations for pure shear-thinning (Carreau-Yasuda) and viscoplastic (Herschel-Bulkley-extended) models. The results of the comparative analysis confirm the reliability and robustness of this generalized Reynolds number in characterizing and interpreting flow behavior in systems with visco-inelastic non-Newtonian fluids. This unified generalized Reynolds number formulation presents new and significant opportunities for precise flow characterization and interpretation as it is applicable to any visco-inelastic (time-independent) viscosity model without requiring additional derivations.

## I. INTRODUCTION

The first flow field characterization at the steady conditions (known as the Reynolds number) was developed by Osborne Reynolds in 1883 to understand when the eddy structures occur in a flow field and how the energy dissipation can be quantified [1]. The properties of Newtonian fluids have been well-studied and established throughout the past 150 years, whereas our knowledge of non-Newtonian fluids is still limited, considering the available viscosity models and the broad range of parameters [2-5]. These fluids experience strong differences in their pressure and stress fields [6, 7] that cannot be characterized with the original definition of the Reynolds number, which is only valid for constant viscosity fluids. It is well established that the varying viscosity of the non-Newtonian fluids often leads to shifts in non-dimensional numbers that are essential to compare and characterize individual flow fields [2, 3, 8].

The real impact of non-Newtonian fluids cannot be isolated, and the flow field characteristics cannot be replicated without the dynamic similarity of dimensionless parameters (known as Π-theorem) [9, 10]. However, many studies did not fully consider these groups where they kept the geometry and flow conditions constant and only changed the fluid rheology [8, 11-13]. In these cases, the corresponding independent dimensionless numbers (Π groups) reduces to only Reynolds number [10]. In a notable study by Gijsen et al.[14], it was shown how non-matching Π groups led to enhanced differences in the flow field. They obtained the velocity profile of a shear-thinning fluid in a 90° curved tube using laser doppler anemometry, and numerically replicated the experimental conditions with a Newtonian fluid by matching the non-dimensional groups, including the Reynolds number. Their results revealed that such matching yields similar bulk flow features (velocity profile and vortical structures) while the wall shear stress and vortex dynamics (vorticity, vortex size and location) are different [14]. Although the results presented by Gijsen et al. [14] were limited, they highlight the importance of matching the Π groups for a physically accurate interpretation of non-Newtonian effects. Such a study shows that the Π group treatment isolates shear-dependent effects and significantly changes the flow dynamics. This raise the concern that previous study results might be biased due to the mismatched Reynolds numbers between systems. The first step toward addressing this issue is to use generalized Reynolds number formulation[2]. However, all currently available generalized Reynolds numbers are viscosity-model-dependent[2, 3, 15]. This highlights the need for the development of a model independent generalized Reynold number.

Since the original definition (and the most commonly used) of Reynolds number is only valid for constant viscosity fluids, it is not suitable to characterize the flow of non-Newtonian fluids[2]. Such characterizations require the Reynolds number formulation to be modified (so-called *generalized*) for non-Newtonian viscosity models. The first generalization of Reynolds number was introduced by Metzner and Reed in 1955 to quantify the frictional losses and the flow transition point of power-law fluids[2]. Following their derivation methodology, Madlener et al.[3] and Dosunmu et al.[15] developed Reynolds number formulations that can characterize the Herschel-Bulkley-extended (HBE) and Carreau fluids, respectively. Although these formulations are called generalized Reynolds numbers, all of them implicitly contain a specific viscosity model in their derivation. Thus, their *generalizability* is strictly limited to their respective viscosity models. In order to overcome the applicability limitation of generalized Reynolds numbers, we are proposing an effective shear rate-based approach for a viscosity-model independent formulation that can be used for any visco-inelastic non-Newtonian fluids, such as blood, polymers, gel fuels. This proposed method adopts the measurement techniques of conventional viscometers to obtain a *unified* Reynolds number formulation. Herein, we experimentally assess the validity of this novel formulation for various viscosity models and parameters by performing friction factor experiments. In addition, we evaluated our proposed model using published experimental data [3, 6] and compared it with established generalized viscosity-model-dependent Reynolds numbers [2, 3, 15].

## II. THEORY

The first step in our derivation for a viscosity-model-independent Reynolds number is to utilize the inverse linear relation between the *apparent* (true) Reynolds number ($Re_{app}$) and Darcy friction factor ($f$) in the laminar regime[1]. Such linear relation exists for any fluid when the Reynolds number is generalized [1-3, 15]. For a straight tube (or channel) this relation can be written as[16],

$$Re_{app} = \frac{64}{f} \qquad (1)$$

The Darcy friction factor can be expressed either in terms of wall shear stress or the pressure gradient using the well-known Darcy–Weisbach equation [17] and force balance ($\tau_{wss} = \Delta p D / 4L$) as,

$$f = \frac{8}{\rho \bar{u}^2} \tau_{wss} = -\frac{2D}{\rho \bar{u}^2} \frac{\Delta p}{L} \qquad (2)$$

where $\tau_{wss}$ is the wall shear stress, $\bar{u}$ is the average velocity over the cross section, $\Delta p/L$ is the pressure gradient and $D$ is the hydraulic diameter of the tube (or channel). Equations 1 and 2 can be combined to obtain the *apparent* (true) Reynolds number as,

$$Re_{app} = \frac{8\rho \bar{u}^2}{\tau_{wss}} \qquad (3)$$

The wall shear stress can also be expressed as a function of dynamic viscosity, $\mu$, and the shear rate at the wall, $\dot{\gamma}_w$.

$$\tau_{wss} = \mu \dot{\gamma}_w \qquad (4)$$

This wall shear stress equation can be extended to a visco-inelastic fluid by considering its local viscosity at the wall as,

$$\tau_{wss} = \mu(\dot{\gamma}_w)\, \dot{\gamma}_w \qquad (5)$$

From equation 5 it is evident that the generalization of Reynolds number requires a surrogate for the wall shear rate, $\dot{\gamma}_w$. Some surrogate formulations have been developed by Mooney[18] and Rabinowitsch[19] by including the fluid viscosity model to consider the slip and fluidity effects of fluids. Using these surrogates requires viscosity model assumption in the intermediate step of the derivation, that leads to a rigorous derivation, and couples the viscosity to the wall shear rate. Thus, the previously developed formulations contain the viscosity model implicitly and are only applicable to their respective viscosity model.

In order to overcome the limitation associated with the viscosity model dependency, we are proposing to directly adapt the measurement principles of rotational viscometers to estimate the wall shear rate [20, 21]. Such viscometers assume the shear rate along the measurement region is uniform (irrespective to the tested fluid), and the wall shear stress is proportional to this shear rate[21]. Then, the viscosity parameters are obtained by curve fitting to such experimental measurements where these assumptions are inherent[22, 23]. Herein, we suggest this assumption of the measurement devices can also be applied for the wall shear rate calculation. We hypothesize the accuracy of Reynolds number formulation would not be affected by this assumption since it is already included to obtain empirical viscosity parameters. Thus, we incorporate the aforementioned assumptions to the Mooney's wall shear rate expression (consult [18] for further details) and simplify it as the following. (Equation 6).

$$\dot{\gamma}_{wN} = \frac{8\bar{u}}{D} \qquad (6)$$

We can also extend this shear rate definition to estimate the effective wall shear stress and viscosity of a flow field (Equation 5). Then, our proposed unified Reynolds number, $Re_{eff}$, can be summarized as,

$$Re_{eff} = \frac{8\rho \bar{u}^2}{\tau_{wss}} = \frac{8\rho \bar{u}^2}{\mu(\dot{\gamma}_{wN})\, \dot{\gamma}_{wN}} \qquad (7)$$

This proposed Reynolds number formula (equation 7) reduces to the conventional Reynolds number definition $Re_{eff} = \rho \bar{u} D / \mu$ for a Newtonian fluid ($\mu = const.$), and it can be directly applied to any viscosity formulation without additional derivation.

For a Carreau-Yasuda (CY) fluid, the proposed formulation can be adapted to,

$$Re_{CY} = \frac{8\rho \bar{u}^2}{(\mu_0 - \mu_\infty) \frac{8\bar{u}}{D} \left[1 + \left(\lambda_c \frac{8\bar{u}}{D}\right)^a\right]^{\frac{n-1}{a}} + \mu_\infty \frac{8\bar{u}}{D}} \qquad (8)$$

where $\mu_\infty$ is the infinite shear rate viscosity, $\mu_0$ is the zero-shear rate viscosity, $\lambda_c$ is the Carreau time constant, $n$ is the power index, and $a$ is the transition parameter from the zero-shear rate region to the power law region.

Additionally, the proposed formulation can easily be applied to a viscoplastic viscosity model (models with yield stress) without any additional modification. For a viscoplastic model example, the following equation can be written for the Herschel-Bulkley-extended (HBE) viscosity model,

$$Re_{HBE} = \frac{8\rho \bar{u}^2}{\tau_0 + K_{HBE}\left(\frac{8\bar{u}}{D}\right)^{n_{HBE}} + \mu_\infty \frac{8\bar{u}}{D}} \qquad (9)$$

where $\tau_0$ is the yield stress of the fluid, $K_{HBE}$ is the HBE power factor, and $n_{HBE}$ is the HBE power exponent.

The previously developed models cannot be applied to multiple viscosity models due to their implicit viscosity model inclusion. The above viscosity model examples, the CY and HBE viscosity models, are consciously selected as their generalized Reynolds number formulations are available in the literature (See Supplementary Information)[3, 15]. Thus, the validity of our proposed method can be assessed by comparing against these formulations.

### III. EXPERIMENTAL METHODS AND MATERIALS

Friction factor experiments were performed in order to test our formulation for power-law and Carreau-Yasuda models with various viscosity parameters. The experimental setup is described in section 3.1, test fluids are presented in section 3.2, and the measurement equipment are introduced in section 3.3.

#### A. Experimental setup

The experimental setup for the friction factor measurements is shown in figure 1. These measurements were taken in a straight, rigid glass tube with a diameter of 5.46 mm (Black lines in *figure 1*). The glass tube was customized with barbed ends and adaptor openings in the middle section. The latter allows pressure measurement catheters to be directly inserted to the measurement location in the test section.

The flow was supplied with one of the two distinct kind of pumps based on desired flow rate. The low flowrate conditions ($\leq 400 ml/min$) were generated with a syringe pump (PHD Ultra, Harvard Apparatus, MA). The higher flowrate values ($> 400\ ml/min$) were generated by an rpm-controllable centrifugal pump (Cancion, Orqis Medical Coorp, CA). These pumps operate on an independent flow circuit that was controlled by a three-way valve in between the pumps and test section.

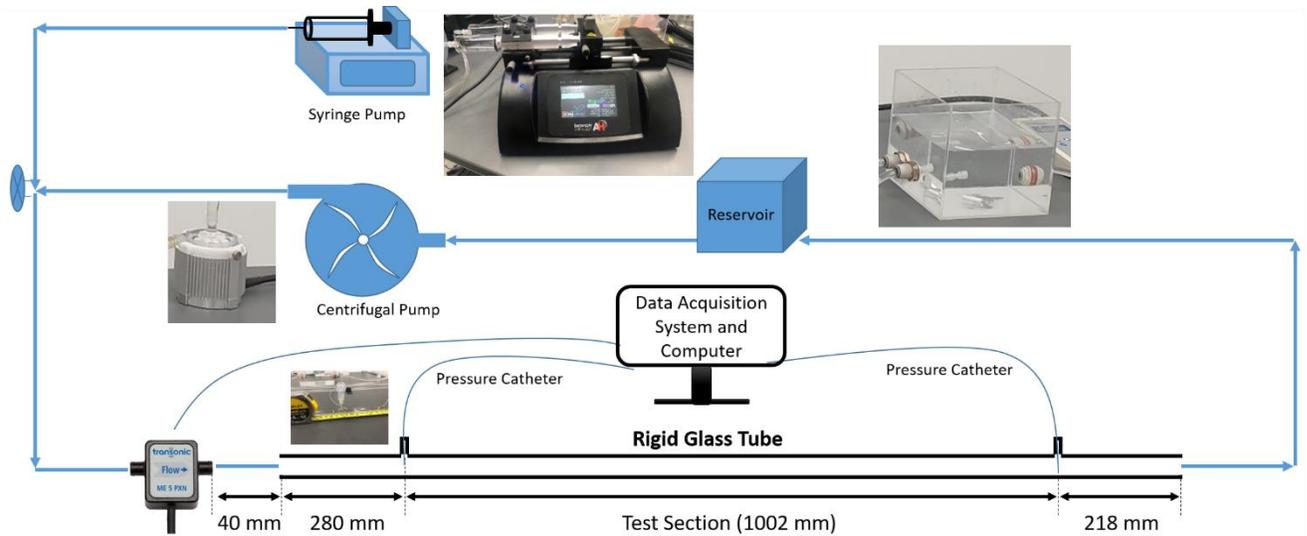

Figure 1: Illustration of the experimental setup for friction factor measurements. The test section (depicted by black lines) is a straight, rigid glass tube. The flow was supplied by one of the two different types of pumps. These pumps operated on an independent flow circuit, which was regulated by a three-way valve positioned between the pumps and the test section.

### B. Materials - Testing Fluids

Total of five test fluids were used in the experiments, consisting of two Newtonian fluids and three non-Newtonian fluids. The Newtonian fluids, deionized water (DIW) and a mixture of glycerol-DIW (45-55 vol.%), were used to test the capability of the experimental setup to capture friction factor. In order to test our Reynolds number formulation for a range of viscosity parameters, three aqueous solutions of Xanthan gum (XG) from Sigma-Aldrich (Tseinheim, Germany) were prepared. Two of these XG solutions were prepared with weight percentages of 0.04% and 0.05%, and the other solution was prepared without weighing the XG amount for the blind tests. These solutions will be referred as XG1, XG2, and XG3 for the 0.04wt.% and 0.05wt.% and unscaled mixtures, respectively.

Before every experiment, the testing fluid was stirred in a closed container for at least 12 hours using a magnetic stirrer. The fluid was then transferred to the experimental setup and circulated for an hour with the centrifugal pump to ensure its homogeneity before the measurements.

After the data collection, a small amount of fluid ($\sim 20 mL$) was drawn from the reservoir for viscosity measurements. These measurements were collected with a rotating viscometer (IKA ROTAVISC lo-vi advanced, Seoul, Korea), and co-centric cylindrical spindles (DINS-1, IKA ROTAVISC). The spindles are placed in a water bath to ensure the viscosity measurements were performed at the same temperature as the flow measurements. The viscosity values of the Newtonian fluids (DIW and glycerol mixture) were measured at two shear rates ($1s^{-1}$ and $100s^{-1}$) to avoid systematic errors. For the aqueous XG mixtures, the viscosities were measured in the shear rate range of (0.1-250 $s^{-1}$) by gradually increasing the shear rate. A MATLAB script was used to find their corresponding Carreau-Yasuda and Power-Law parameters. These viscosity parameters are tabulated in *Table I*. *Figure 2* shows an example of the fitted Carreau-Yasuda and power-law curves to the viscosity measurements of XG3 (See Supplementary Information for the measurements of XG1 and XG2).

TABLE I: Viscosity parameters and densities of the fluids used in the experiments

| | | Carreau-Yasuda Model | | | | Power-Law Model | | Density |
|---|---|---|---|---|---|---|---|---|
| | $\mu_\infty (mPa.s)$ | $\mu_0 (mPa.s)$ | $\lambda_c (s)$ | $n$ | $a$ | $K\ (mPa.s^{n_{PL}})$ | $n_{PL}$ | $\rho\ (kg/m^3)$ |
| XG1 | 1.01 | 71.1 | 1.198 | 0.4733 | 1.877 | 54.92 | 0.567 | 997 |
| XG2 | 1.01 | 124.0 | 2.213 | 0.4495 | 1.921 | 69.94 | 0.522 | 998 |
| XG3 | 1.01 | 196.2 | 2.835 | 0.363 | 1.944 | 73.42 | 0.515 | 999 |
| DIW | 1.01 | - | - | - | - | - | - | 997 |
| Glycerol + DIW (45 vol.%) | 5.61 | - | - | - | - | - | - | 1127 |

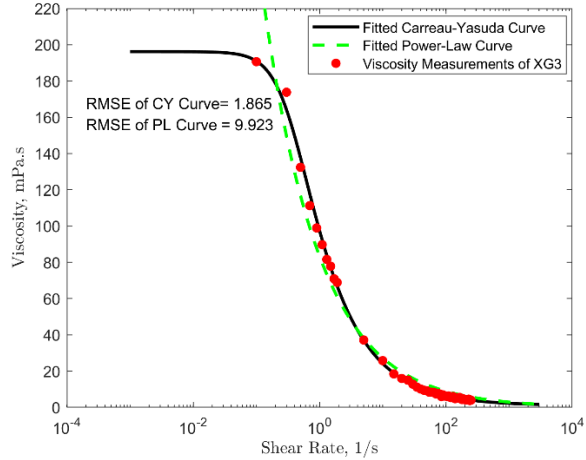

Figure 2: The measured viscosity values for XG3, and its corresponding (fitted) Carreau-Yasuda and power-law curves.

### C. Data Collection Devices

The flow rate values were measured using an inline ultrasonic flowmeter (ME-6PXN, Transonic Systems Inc.), which was placed 140 mm before the inlet of the glass tube. The pressure data were acquired using high-fidelity pressure catheters (Mikro-Cath, Millar Inc., Houston, TX) that were inserted directly into the testing section through the adaptor ports. To minimize the entrance and end effects on the measurements, the catheters were positioned at distances of 280 mm from the inlet and 218 mm from the outlet of the testing tube. The data from these sensors were acquired using a PowerLab data acquisition system (ADInstruments, Colorado Springs, CO) and its bundled software, LabChart Pro 7.

### D. The robustness of the experimental setup

Two Newtonian fluids (deionized water (DIW) and 45 vol.% Glycerol) are used to assess our experimental setup's capability. In figure 3, the Reynolds-Darcy relation ($f = 64/Re_{app}$) is shown with a solid line, and the measurements from the experiments are shown with squares and circles for DIW and 45 vol.% Glycerol, respectively. The experimental measurements show a good agreement with the analytical Darcy-Reynolds friction factor relation with the Pearson correlation coefficient ($R^2$) of 0.950, and the root-mean-square error ($RMSE$) is 0.065 up to Reynolds number of 2,000. Thus, this experimental setup is suitable to obtain the friction factors by measuring the pressure drop and flow rate, and the flow transition to turbulence starts at $Re = 2,000$ in the test section.

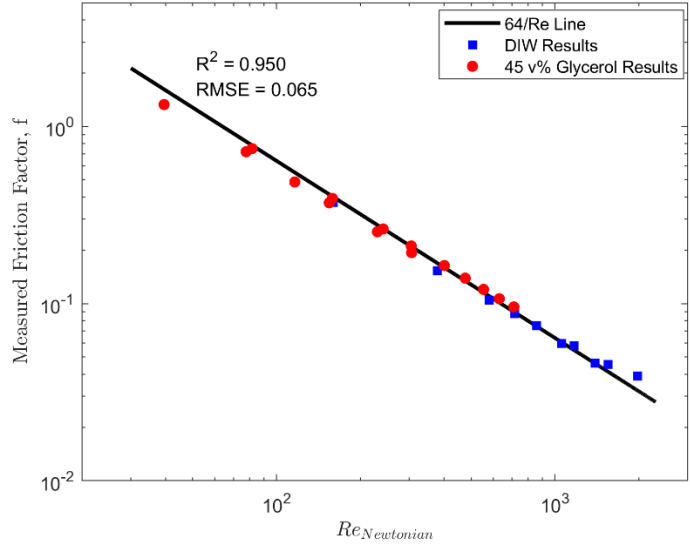

Figure 3: Validation of experimental setup using Newtonian fluids. The experimental measurements (Markers) show good agreement ($R^2 = 0.950, RMS = 0.065$) with the analytical Reynolds-Darcy relation (Black line) up to $Re = 2,000$, confirming the reliability of the setup for friction factor determination through pressure drop and flow rate measurements.

## IV. RESULTS AND DISCUSSION
### *A. The friction factor experiments to assess the proposed formulation*

Figure 4 demonstrates our experimental results overlaid with the estimations of our proposed formulation and the established *generalized* (viscosity-model-dependent) Reynolds number formulation for the power-law model[2]. The values on the horizontal axis are the Reynolds number predictions of our proposed model, and Metzner and Reed's estimations are adjusted accordingly. The proposed Reynolds number formulation accurately predicts the experimental data ($R^2$ of $0.982$, the RMSE is $0.149$). Whereas Metzner and Reed's formulation has a constant bias of 13% from our predictions, as well as the experimental data. A similar overestimation of the true Reynolds number by Metzner and Reed's formulation can also be observed in their original publication (Figure 1A of Metzner and Reed[2]), where they introduced the generalized Reynolds number formulation for the power-law fluids[2]. Such a constant shift between the predictions should be attributed to the theoretical limits of power-law viscosity model as it does not have the viscosity plateaus ($\mu \rightarrow 0$ as $\dot{\gamma} \rightarrow \infty$, vice versa). Therefore, the contribution of the wall shear rate surrogates is linearly proportional in the Reynolds number formulations. Despite the deviation observed for the power-law viscosity model, our proposed model demonstrates superior accuracy in estimating the experimental data compared to the established formulation. Consequently, we anticipate that our approach will successfully characterize the flow physics of power-law fluids.

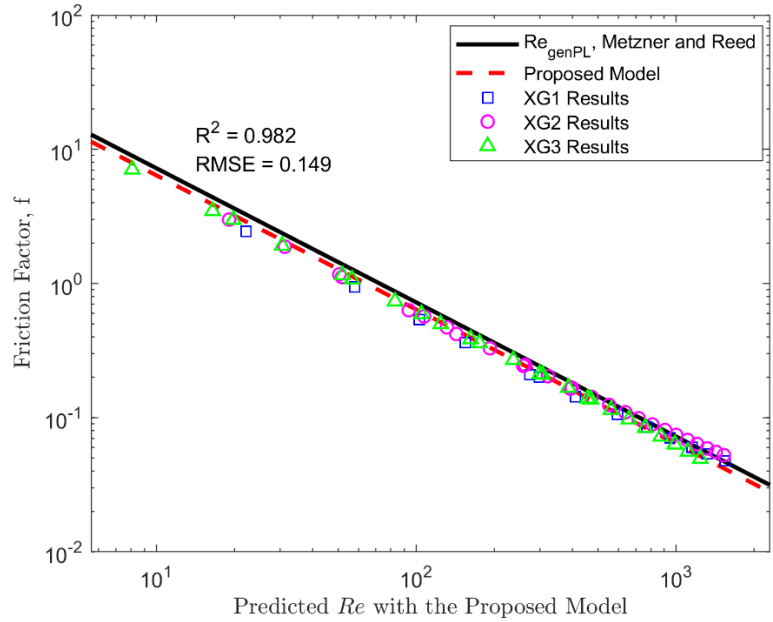

Figure 4: Experimental assessment of the proposed generalized formulation (Red dashed line) against the (modified) generalized Reynolds number for power-law model fluids (Black solid line). The results show the correlation factor, $R^2$, of 0.982 and $RMSE\ of$ 0.149 between our model and the experimental results. The predictions of Metzner and Reed's formulation has $R^2\ of$ 0.904 $and\ RMSE\ of$ 0.345 with the experimental data. The constant bias of 13% of Metzner and Reed's formula results in a lower accuracy in the predictions of the experimental data.

In order to test our method with a relatively more complex viscosity model, we proceed our evaluations with the Carreau-Yasuda viscosity model. To compare our model with an established Reynolds number in figure 5, we modified the formulation of Dosunmu et al.[15] for the Carreau-Yasuda model. Figure 5 shows the collected experimental data compared against the predictions of both our proposed formula and the modified model of Dosunmu et al.[15]. The Reynolds number values on the horizontal axis are determined using the estimations of our proposed model, and the friction factor predictions of Dosunmu et al.[15] are adjusted accordingly. Whereas, the Newtonian approximation line is drawn by estimating the friction factors with the infinity shear rate viscosity value of the fluids, ($Re_{Newtonian} = \rho \bar{u} D / \mu_\infty$), this Reynolds number definition has been used for non-Newtonian fluids in the literature [3, 11, 24].

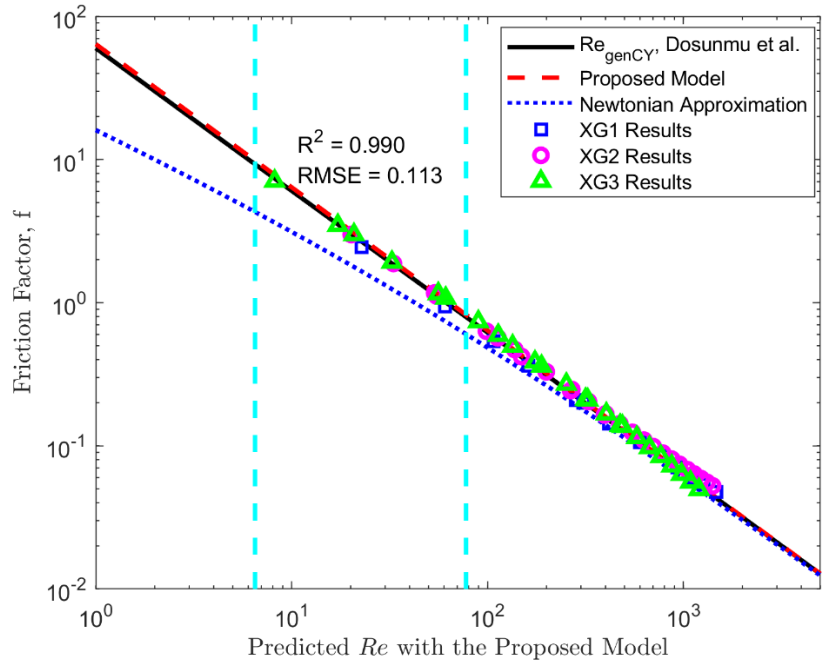

Figure 5: Experimental comparison of the predictions of the proposed generalized formulation (Red dashed line) against the (modified) generalized Reynolds number for Carreau-Yasuda fluids (Black solid line). The proposed model's predictions have the correlation factor, $R^2$, of 0.990 and $RMSE\ of\ 0.113$ with the experimental results. The Newtonian Reynolds number approximation (blue dotted lines) has $R^2\ of\ 0.905\ and\ RMSE\ of\ 0.343$ with the experiments.

In order to demonstrate the mispredictions of the Newtonian approximation, we mark two Reynolds numbers on figure 5 (with dashed vertical lines) that are equivalent to the common biological shear rate values (14-100 s$^{-1}$) [11, 12]. The lower shear-rate value (14 s$^{-1}$) has been observed in multiple locations of human body [8, 12, 25], and the latter value (100 s$^{-1}$) is the commonly accepted shear thinning limit of the human blood [25, 26]. Within this region, we observe significant differences between the predictions of the generalized Reynolds numbers and Newtonian approximation (128%-36%). This non-negligible difference of the Newtonian approximation might be misleading in predicting the flow physics.

The previous shift observed for power-law fluids between the curves of (true) Reynolds number predictions diminish as more viscosity parameters are included for the Carreu-Yasuda than power-law fluids. The maximum difference between these predictions occur at a low Reynolds number ($Re\sim 0.01$) as 6.55%. Both the proposed and established Reynolds number formulations show great agreement with the friction factor experiments ($R^2 = 0.990, RMS = 0.113$), and they can estimate the flow physics by accurately predicting the true Reynolds number.

### B. Comparative analysis of the proposed generalized Reynolds number against the established Reynolds numbers for HBE and Carreau fluids

Our proposed model is compared against the viscosity specific Reynolds number formulations for HBE and Carreau fluids figures 6.a and 6.b, respectively. The Reynolds number curve for the HBE fluid (Figure 3.a) is drawn using the rheology parameters of 85 wt.% kerosene, 7.5 wt.% thixatrol ST and 7.5 wt.% miak solution[3]. The curve of the Carreau fluid (Figure 3.b) is drawn using the parameters for the whole human blood [27]. Whereas, the "Newtonian" Reynolds number, $Re_{Newtonian} = \rho \bar{u} D / \mu_\infty$, lines are drawn using infinity shear rate viscosity values of the fluids, such Reynolds number definition has been used in the literature [3, 11, 24].

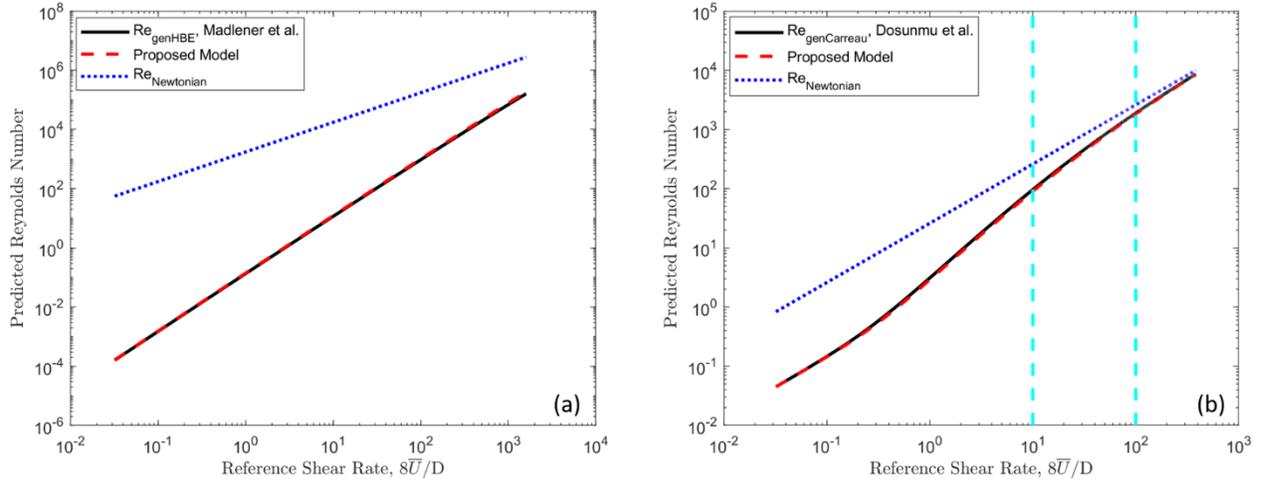

Figure 6: The proposed generalized Reynolds number (Red dashed line) compared against the established Reynolds number formulations (Black solid line) and the Newtonian approximation (Blue dotted line) to characterize the flow physics, using (a) an HBE fluid, (b) a Carreau fluid.

The maximum difference between the new model and the generalized formulation of the HBE fluid is observed as 12.8% at the shear rate of 1,000. Whereas, the Carreau fluid shows a maximum deviation of 5.5% between the predictions at Reynolds number of 45. Thus, our proposed approach displays great agreements with two distinct generalized Reynolds number formulations. This shows our formulation can accurately estimate the flow physics regardless of the viscosity model.

In figure 6.a, the Newtonian approximation shows considerable overestimation for the Reynolds number estimations for the HBE fluid compared to both our and established generalized Reynolds numbers. At low shear rates, $\dot{\gamma} \approx 1 s^{-1}$, this overestimation is more than $10^4$ times of the true Reynolds number, which would lead to severe differences in the flow characterization. Thus, introducing potential errors to the flow analysis and observations. In the context of kerosene (a gelled fuel), mischaracterizing the flow regime may have implications for the design and performance of fuel injection systems and combustion chambers [28-30]. Such an incorrectly characterized flow field may lead to inaccurate predictions on fuel mixing and combustion efficiency[28, 31].

For the Carreau fluid, the Newtonian assumption has a significant deviation from the true Reynolds number curves. In order to quantify the significance of these deviations we highlight the common biological shear rate values (similar to figure 6) with two dashed vertical lines on figure 7.b. Within this physiological shear rate region, the Newtonian approximation may overestimate the true Reynolds number up to 138%. Although this difference appears to be decreasing by increased shear rates, the Newtonian assumption still differs from the true Reynolds number by 25% at higher shear rate values, $\dot{\gamma} > 200 \, s^{-1}$. As the Newtonian Reynolds number approximations show significant differences for both HBE and Carreau fluids, using this definition would be misleading while comparing to systems and characterizing the flow physics. Such importance of correct flow characterization for comparing two individual systems was shown in the preliminary results of Gijsen et al.[14]. In order to address these challenges and isolate the effects of viscosity, our proposed formulation can be utilized to estimate the true Reynolds number of different systems, facilitating a more precise characterization of flow physics and better understanding of viscosity-related flow phenomena.

### C. Evaluation and validation of the proposed Reynolds number formulation using available pressure drop data

In order to further test the accuracy of our proposed formulation, we evaluate our model's predictions using the already available friction factor (pressure drop) experiments of HBE[3] and Carreau fluids[6]. Figure 7.a shows the comparison of the experimental data and the predictions of the proposed and established Reynolds number formulations for an HBE fluid. Figure 7.b presents the experimental data of the Carreau fluid and the Reynolds number estimations. The horizontal axis of these plots (Reynolds number values) are determined by the generalized (viscosity-model-dependent) Reynolds number formulations (in accordance with the original publications of the experiments), and the predictions of our model are adjusted accordingly.

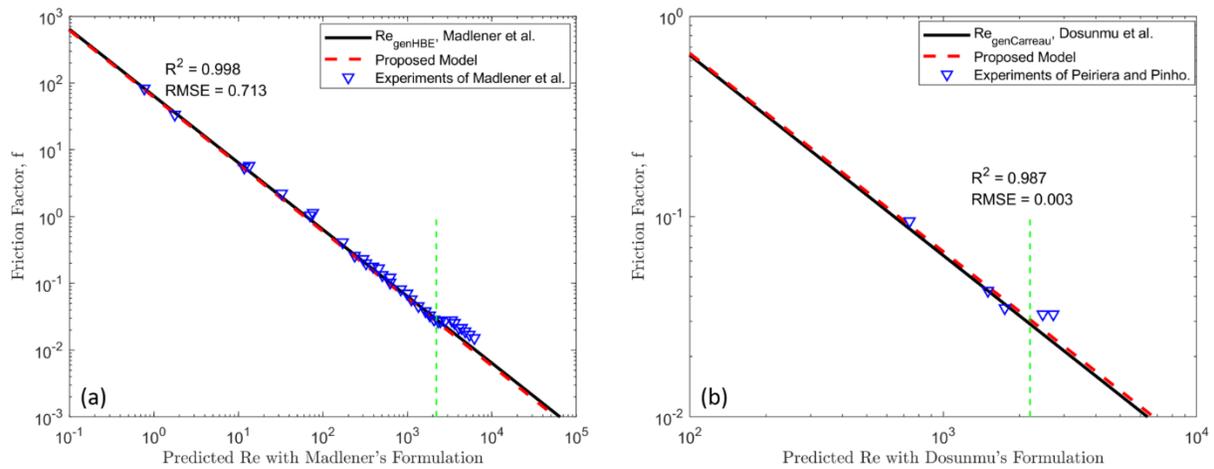

Figure 7: The proposed generalized Reynolds number formulation (red dashed line) is compared against previously published experimental data (markers) and the established generalized Reynolds number formulations (Black solid line) for (a) HBE fluid and (b) Carreau fluid. The green line depicts the limit of the laminar flow regime, $Re = 2,200$. The proposed model's predictions show great agreement ($R^2 > 0.98$) with these experimental data in the laminar flow regime.

In the laminar region ($Re < 2,200$, annotated in the figures), our approach shows similar trends and values to Madlener et al.[3]'s formulation for the HBE fluid, and both methods are in great agreement with the experimental data (Figure 7.a). The greatest difference between our predictions and the experimental results is observed at Reynolds number of 76 with 32.4%, whereas the established (viscosity-model-dependent) Reynolds number deviates by 26.5%. Although the Carreau fluid (Figure 4.b) has limited data[6] in the laminar region, our proposed formulation successfully estimates the true Reynolds numbers with a maximum difference of 8.5% at Reynolds number of 747.

Our proposed Reynolds number formulation shows excellent agreement with various viscosity models and parameters, when compared with our experiment results, as well as the already established viscosity-model-dependent Reynolds number formulas. We claim our formulation can be extended to any visco-inelastic fluid to predict the effective (true) Reynolds number and characterize the flow physics without any additional derivation. This formulation may benefit experimental and computational fluid mechanics investigations. Our proposed method uses an approximation for the effective wall shear rate to calculate the true Reynolds number. Using this approximation, the experimental researchers could be better informed the main flow physics by using commonly available fluids without modifying their setups[9], or working with non-practical (flammable, toxic, etc.) fluids[11, 32]. Although most of the computational investigations are inspired by non-Newtonian fluids (blood, combustion, injection etc.), the original rheology of the fluids is often not fully considered and such simulations may have a different flow physics. Such an issue can be addressed by using our proposed generalized Reynolds number that offers a robust method to estimate the true Reynolds number for any viscosity model. As evident in the study of Gijsen et al[14], in order to isolate the non-Newtonian effects, the flow physics should be precisely characterized and equalized between systems. Our proposed formulation would offer an easily applicable method to quantify the true Reynolds numbers of such systems. Additionally, the viscosity-model independency of our proposed formulation allows it to be extended to viscosity models without an established Reynolds number formulation (such as Quemada[33] and Casson[34] fluids). Overall, the flow characterization of polymeric solutions (viscoelastic fluids) necessitates additional dimensionless numbers (such as Deborah and Weissenberg) other than the Reynolds number. Thus, the proposed flow characterization method is more suitable for time-independent non-Newtonian fluids. The presented experimental results only show the accuracy of the proposed Reynolds number formulation for shear-thinning fluids, while using the polymeric solutions.

## V. LIMITATIONS

The rotational viscometer is capable of accurate viscosity measurement in the shear rate range of $0.01 - 250 s^{-1}$. Therefore, the authors had to extrapolate the zero shear rate viscosity parameters, $\mu_0$, using the available data points. The measurement of this parameter may require a more sensitive device (such as an air-bearing rheometer) that was not available to the authors; however, this won't affect the overall finding of this manuscript. Furthermore, the agreement between the experimental results and Reynolds number formulations (both the established and proposed) indicates that a sensitive measurement of $\mu_0$ was not necessary to prove the suggested concept.

## VI. CONCLUSIONS

In this study, we introduced a new viscosity-model-independent formulation for a unified generalized Reynolds number for visco-inelastic non-Newtonian fluids. We experimentally tested the accuracy of our proposed formulation using power-law and Carreau-Yasuda fluids with different viscosity parameters by conducting friction factor experiments. The results of our experiments demonstrated the applicability and effectiveness of this novel generalized Reynolds number. Furthermore, the accuracy of this Reynolds number is tested against the established generalized Reynolds formulations such as the Carreau-Yasuda (pure shear-thinning) and Herschel-Bulkley-extended (viscoplastic) models. The wide applicability of this model-independent generalized Reynolds number eliminates the need for additional derivations for each viscosity model. This also indicates that our approach provides a reliable characterization of flow dynamics of visco-inelastic fluids. Such a viscosity-model-independent formulation provides an adaptable tool for more accurate comparisons and interpretations of flow dynamics of non-Newtonian fluids.

## REFERENCES

[1]  O. Reynolds, "XXIX. An experimental investigation of the circumstances which determine whether the motion of water shall be direct or sinuous, and of the law of resistance in parallel channels," *Philosophical Transactions of the Royal Society of London,* vol. 174, pp. 935-982, 1883.

[2]  A. Metzner and J. Reed, "Flow of non-newtonian fluids—correlation of the laminar, transition, and turbulent-flow regions," *Aiche journal,* vol. 1, no. 4, pp. 434-440, 1955.

[3]  K. Madlener, B. Frey, and H. Ciezki, "Generalized reynolds number for non-newtonian fluids," *Progress in propulsion physics,* vol. 1, pp. 237-250, 2009.

[4]  T. Shende, V. J. Niasar, and M. Babaei, "Effective viscosity and Reynolds number of non-Newtonian fluids using Meter model," *Rheologica Acta,* vol. 60, no. 1, pp. 11-21, 2021.

[5]  F. Pinho and J. Whitelaw, "Flow of non-Newtonian fluids in a pipe," *Journal of non-newtonian fluid mechanics,* vol. 34, no. 2, pp. 129-144, 1990.

[6]  A. S. Pereira and F. Pinho, "Turbulent pipe flow characteristics of low molecular weight polymer solutions," *Journal of non-newtonian fluid mechanics,* vol. 55, no. 3, pp. 321-344, 1994.

[7]  R. A. Chilton and R. Stainsby, "Pressure loss equations for laminar and turbulent non-Newtonian pipe flow," *Journal of hydraulic engineering,* vol. 124, no. 5, pp. 522-529, 1998.

[8]  C. Bilgi and K. Atalık, "Numerical investigation of the effects of blood rheology and wall elasticity in abdominal aortic aneurysm under pulsatile flow conditions," *Biorheology,* vol. 56, no. 1, pp. 51-71, 2019.

[9]  E. Buckingham, "Physically similar systems," *Journal of the Washington Academy of Sciences,* vol. 4, no. 13, pp. 347-353, 1914.

[10]  A. A. Sonin, "A generalization of the Π-theorem and dimensional analysis," *Proceedings of the National Academy of Sciences,* vol. 101, no. 23, pp. 8525-8526, 2004.

[11]  H. Wei, A. L. Cheng, and N. M. Pahlevan, "On the significance of blood flow shear-rate-dependency in modeling of Fontan hemodynamics," *European Journal of Mechanics-B/Fluids,* vol. 84, pp. 1-14, 2020.

[12]  C. Bilgi and K. Atalık, "Effects of blood viscoelasticity on pulsatile hemodynamics in arterial aneurysms," *Journal of Non-Newtonian Fluid Mechanics,* vol. 279, p. 104263, 2020.

[13]  A. L. Cheng, N. M. Pahlevan, D. G. Rinderknecht, J. C. Wood, and M. Gharib, "Experimental investigation of the effect of non-Newtonian behavior of blood flow in the Fontan circulation," *European Journal of Mechanics-B/Fluids,* vol. 68, pp. 184-192, 2018.

[14]  F. J. H. Gijsen, E. Allanic, F. N. van de Vosse, and J. D. Janssen, "The influence of the non-Newtonian properties of blood on the flow in large arteries: unsteady flow in a 90° curved tube," *Journal of Biomechanics,* vol. 32, no. 7, pp. 705-713, 1999/07/01/ 1999.

[15]  I. T. Dosunmu and S. N. Shah, "Pressure drop predictions for laminar pipe flow of carreau and modified power law fluids," *The Canadian Journal of Chemical Engineering,* vol. 93, no. 5, pp. 929-934, 2015.

[16]  O. Reynolds, "XXIX. An experimental investigation of the circumstances which determine whether the motion of water shall be direct or sinuous, and of the law of resistance in parallel channels," vol. 174, pp. 935-982, 1883.

[17]  H. Darcy, *Recherches expérimentales relatives au mouvement de l'eau dans les tuyaux*. Mallet-Bachelier, 1857.

[18]  M. Mooney, "Explicit formulas for slip and fluidity," *Journal of Rheology (1929-1932),* vol. 2, no. 2, pp. 210-222, 1931.

[19]  B. Rabinowitsch, "Über die viskosität und elastizität von solen," *Zeitschrift für physikalische Chemie,* vol. 145, no. 1, pp. 1-26, 1929.

[20] V. Kelessidis and R. Maglione, "Modeling rheological behavior of bentonite suspensions as Casson and Robertson–Stiff fluids using Newtonian and true shear rates in Couette viscometry," *Powder technology,* vol. 168, no. 3, pp. 134-147, 2006.

[21] P. Estellé, C. Lanos, and A. Perrot, "Processing the Couette viscometry data using a Bingham approximation in shear rate calculation," *Journal of Non-Newtonian Fluid Mechanics,* vol. 154, no. 1, pp. 31-38, 2008.

[22] I. M. Krieger and H. Elrod, "Direct determination of the flow curves of non-Newtonian fluids. II. Shearing rate in the concentric cylinder viscometer," *Journal of applied physics,* vol. 24, no. 2, pp. 134-136, 1953.

[23] I. M. Krieger, "Shear rate in the Couette viscometer," *Transactions of the Society of Rheology,* vol. 12, no. 1, pp. 5-11, 1968.

[24] A. M. Walker, C. R. Johnston, and D. E. Rival, "On the characterization of a non-Newtonian blood analog and its response to pulsatile flow downstream of a simplified stenosis," *Annals of biomedical engineering,* vol. 42, no. 1, pp. 97-109, 2014.

[25] Z. Wei *et al.*, "Non-Newtonian effects on patient-specific modeling of Fontan hemodynamics," *Annals of biomedical engineering,* vol. 48, no. 8, pp. 2204-2217, 2020.

[26] S. Berger and L.-D. Jou, "Flows in stenotic vessels," *Annual review of fluid mechanics,* vol. 32, no. 1, pp. 347-382, 2000.

[27] S. Chien, "Biophysical behavior of red cells in suspensions," *The red blood cell,* vol. 2, pp. 1031-1133, 1975.

[28] S. Yang, T. Hsu, and M. Wu, "Spray combustion characteristics of kerosene/bio-oil part II: Numerical study," *Energy,* vol. 115, pp. 458-467, 2016.

[29] Y. Hardalupas, C. Liu, and J. Whitelaw, "Experiments with disk stabilized kerosene-fuelled flames," *Combustion science and technology,* vol. 97, no. 1-3, pp. 157-191, 1994.

[30] F. Wang, R. Liu, M. Li, J. Yao, and J. Jin, "Kerosene evaporation rate in high temperature air stationary and convective environment," *Fuel,* vol. 211, pp. 582-590, 2018.

[31] V. Shirsat and A. Gupta, "Performance characteristics of methanol and kerosene fuelled meso-scale heat-recirculating combustors," *Applied energy,* vol. 88, no. 12, pp. 5069-5082, 2011.

[32] O. K. Baskurt and H. J. Meiselman, "Blood rheology and hemodynamics," in *Seminars in thrombosis and hemostasis*, 2003, vol. 29, no. 05, pp. 435-450: Copyright© 2003 by Thieme Medical Publishers, Inc., 333 Seventh Avenue, New ….

[33] D. Quemada, "Rheology of concentrated disperse systems II. A model for non-newtonian shear viscosity in steady flows," *Rheologica Acta,* vol. 17, no. 6, pp. 632-642, 1978.

[34] L. Srivastava and V. Srivastava, "Peristaltic transport of blood: Casson model—II," *Journal of Biomechanics,* vol. 17, no. 11, pp. 821-829, 1984.

**SUPPLEMENTARY INFORMATION**

1. **The available *generalized* Reynolds number formulations in the literature**

The published generalized Reynolds number formulations can be summarized as:

Suitable for Power-Law model, Metzner and Reed[2]:

$$Re_{gen,PL} = \frac{\rho \bar{u}^{\{2-n\}} D^n}{8^{n-1} K((3n+1)/4n)^n}$$

Suitable for Herschel-Bulkley-Extended model, Madlener et al.[3]:

$$Re_{gen,HBE} = \frac{\rho \bar{u}^{\{-2-n\}} D^n}{\left(\begin{array}{c}\frac{\tau_0}{8}\left(\frac{D}{\bar{u}}\right)^n + K_{HBE}\left(\frac{3m+1}{4m}\right)^n 8^{n-1} \\ +\mu_\infty \frac{3m+1}{4m}\left(\frac{D}{\bar{u}}\right)^{n-1}\end{array}\right)}$$

where, $$m = \frac{nK_{HBE}(8\bar{u}/D)^n + \mu_\infty(8\bar{u}/D)}{\tau_0 + K_{HBE}(8\bar{u}/D)^n + \mu_\infty(8\bar{u}/D)}$$

Suitable for Carreau/Carreau-Yasuda viscosity models, Dosunmu et al.[15]:

$$Re_{gen,CY} = \frac{\rho \bar{u} D}{\left\{\begin{array}{c}\left[\frac{(\mu_0 - \mu_\infty)\left(\frac{3n'+1}{4n'}\right)}{\left(1 + \left(\lambda_c \left(\frac{3n'+1}{4n'}\right)\frac{8\bar{u}}{D}\right)^a\right)^{\frac{1-n}{a}}} + \mu_\infty\left(\frac{3n'+1}{4n'}\right)\right] \\ +(\mu_0 - \mu_\infty)\left[1 + \left(\lambda_c \frac{8\bar{u}}{D}\right)^a\right]^{\frac{n-1}{a}}\end{array}\right\}}$$

where, $$n' = \frac{\frac{8\bar{u}}{D}\left[\begin{array}{c}\mu_\infty - 2\lambda_c^a(\mu_0 - \mu_\infty)\frac{1-n}{a}\left(\frac{8\bar{u}}{D}\right)^a\left\{1 + \left(\lambda_c \frac{8\bar{u}}{D}\right)^a\right\}^{-\left(\frac{n-1}{a}\right)} \\ +(\mu_0 - \mu_\infty)\left\{1 + \left(\lambda_c \frac{8\bar{u}}{D}\right)^a\right\}^{\left(\frac{n-1}{a}\right)}\end{array}\right]}{\mu_\infty \frac{8\bar{u}}{D} + (\mu_0 - \mu_\infty)\frac{8\bar{u}}{D}\left[1 + \left(\lambda_c \frac{8\bar{u}}{D}\right)^a\right]^{\frac{n-1}{a}}}$$

2. **The viscosity measurements of the testing fluids**

Figure S1 shows the viscosity measurements of the testing fluids XG1 and XG2 and their fitted curves for the Carreau-Yasuda and power-law models.

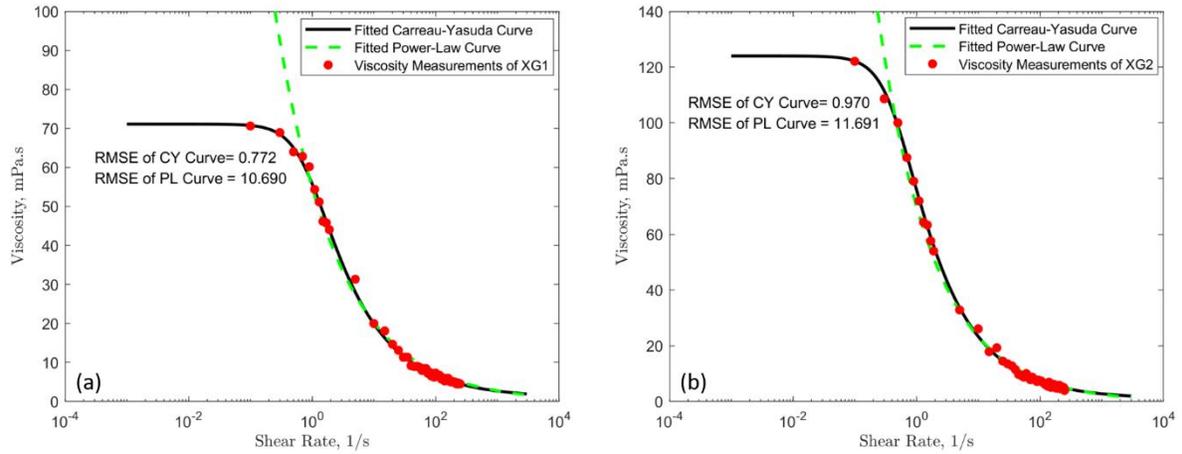

Figure S1. The measured viscosity values overlaid on the fitted Carreau-Yasuda and power-law viscosity curves of the testing fluids of (a) XG1, (b) XG2.